\newif\ifAMStwofonts
\AMStwofontstrue

\def\fg{{f_{_{\rm g} }}}

\def\sigmatwo{{\sigma_{_{2}}}}

\def\gsim{~\rlap{$>$}{\lower 1.0ex\hbox{$\sim$}}}

\def\simpropto{\lower.2ex\hbox{$\; \buildrel \propto \over \sim \;$}}
\def\ltsim{\lower.5ex\hbox{$\; \buildrel < \over \sim \;$}}
\def\gtsim{\lower.5ex\hbox{$\; \buildrel > \over \sim \;$}}
\def\ltsim{\lower.5ex\hbox{$\; \buildrel < \over \sim \;$}}
\def\gtsim{\lower.5ex\hbox{$\; \buildrel > \over \sim \;$}}

\def\Rv{R_{\rm v}}

\def\kms{\mbox{km\,s$^{-1}$}}
\newcommand{\Mpc}{\, {\rm Mpc} }
\def\Msol{\mbox{M$_\odot$}}
\def\mg{ M_{{ g}} }
\def\mbh{ M_{{BH}} }
\def\mbht{ M_{{ BH,8 }} }

\def\mp{ m_{_{\rm p}} }
\def\sigT{ \sigma_{_{\rm T}} }

\def\dd{\,{\rm d}}

\newcommand{\beq}{\begin{equation}}
\newcommand{\eeq}{\end{equation}}
\def\beqa{\begin{eqnarray}}
\def\eeqa{\end{eqnarray}}
\def\fixit#1{}

\def\dd{{\rm d}}

\documentclass[preprint2]{aastex}
\usepackage{natbib}
\shorttitle{The massive black hole-velocity dispersion relation and the halo baryon fraction}
\shortauthors{Nusser \& Silk}

\begin{document}

\title{The massive black hole-velocity dispersion relation and the halo baryon fraction:
a case for positive AGN feedback}
\author{Joseph Silk\altaffilmark{1}}
\affil{Beecroft Institute of Particle Astrophysics and Cosmology, 
Department of Physics, University of Oxford, \\
Denys Wilkinson Building
1 Keble Road
Oxford, OX1 3RH\\
Institut d'Astrophysique de Paris, CNRS, 98bis bd Arago, 75014 Paris, France}
\author{Adi Nusser\altaffilmark{2}}
\affil{ Physics Department and the Asher Space Science Institute, 
Technion, Haifa 32000, Israel}

\altaffiltext{1}{E-mail: silk@astro.ox.ac.uk}
\altaffiltext{2}{E-mail: adi@physics.technion.ac.il}


\begin{abstract}
Force balance considerations put a limit on the rate of  AGN radiation 
momentum  output, $L/c$, capable of driving galactic  superwinds and reproducing the 
observed $\mbh -\sigma $ relation between black hole mass and spheroid velocity dispersion.
We show that  black holes  cannot supply enough momentum
in radiation to drive the gas out by pressure alone. Energy-driven winds give a  $\mbh -\sigma $ scaling favoured by a recent analysis but also fall short energetically once cooling is incorporated.
We propose that outflow-triggering of star formation by enhancing the intercloud medium turbulent pressure and squeezing clouds can supply the necessary boost, and  suggest possible tests of this hypothesis. Our hypothesis simultaneously can account for the observed halo baryon fraction.
\end{abstract}

\section{Introduction}  

\begin{figure}
\epsscale{.8}
 \plotone{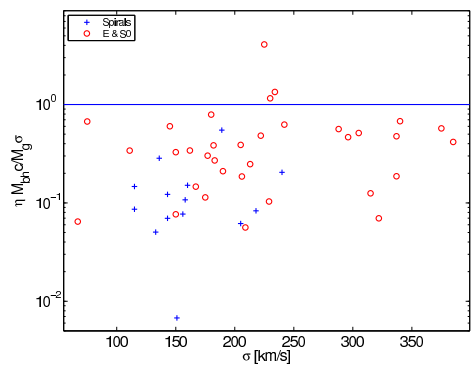}
\caption{Plot of $\eta \mbh c/(\mg \sigma c)$ (for $\eta=0.1$)  versus $\sigma$ from the sample of Gultekin et al 09.
The horizontal line  is 
$0.1\mbh=\mg \sigma /c.$
Black hole momentum falls short of the required momentum for most galaxies. }
\label{fig1}
 \end{figure} 
There is a consensus that  the powerful  Active Galactic Nuclei (AGNs)  play a crucial role in 
shaping the general properties of  galaxies \citep[e.g.,][]{SR,Croton,SN09} and clusters of galaxies \citep[e.g.,][]{Vecc04,NSB,MN07}.
AGNs are powered by accretion onto supermassive black holes believed to reside at the centers of most galaxies. 
An indication of the galaxy-black hole connection is the remarkable correlation between the 
black hole  mass, $\mbh$, and the velocity dispersion, $\sigma$, of the spheroidal galactic components \citep[e.g.,][]{gult09}. 
Any successful model for galaxy formation must provide an explanation of this correlation. Self-regulated black hole growth offers a natural explanation for this relation \citep{SR}. 
Both radiation pressure and mechanical outflows deposit momentum into the protogalactic gas.  If this results in a wind, 
force balance arguments (\citep{Fabian99,King03,Murray,Thompson05} but see \citep{Soker09}) lead to the  conclusion that winds driven by 
pressure of radiation from a central black hole can suppress the collapse of gas and hence 
 regulate the growth of the black hole. However the available momentum is, we show, insufficient to give the required normalization of the $\mbh-\sigma$ scaling (Section 2).

The original self-regulation argument of \cite{SR} relied on energy balance: AGN activity heats  the 
galactic  gas reservoir above the virial temperature, generating galactic winds and eventually terminating gas accretion onto the black hole.     
However, energy-driven winds suffer strong radiative cooling losses:
while the  radiation  heats the gas nearby the black hole, the gas expands but cools rapidly, making the process inefficient (Section 3).
Our preferred solution is to introduce positive AGN feedback via triggered star formation. We argue that this simultaneously resolves three problems: the required order-of-magnitude boost in the $\mbh-\sigma$ scaling
(Section 4), the enhanced specific star formation rate 
in massive galaxies (addressed elsewhere by  \cite{sadegh}, and 
the shortfall in the halo baryon fraction (Section 5).

\section{Can radiation momentum-driven winds expel halo gas?}

A luminosity 
$L/c= \mg g(r)$ balances  the gravitational 
force applied on  gas mass $\mg$ by the dominant dark matter (DM)
in a protogalactic halo. 
For isothermal  DM and gas distributions we get $g(r)=2  \sigma^2 /r$ and
 $\mg(r)=\fg r^2 g(r)/G$ where $\sigma $ is the velocity dispersion and 
 $\fg$  is the gas mass fraction. The force balance condition, 
 with the simplifying assumption  that the gas mass $\mg(r)$ lies entirely on the shell \citep{Murray}, 
 $r$,   yields a minimal luminosity
\begin{equation}
L=\frac{4\fg c \sigma^4}{G}\; .
\end{equation}

If we assume that $ L $ is proportional to $\mbh$ as for
the Eddington luminosity,  $L_{_{\rm Edd}}=4\pi c G \mbh \mp/\sigT$,
this condition translates to
 \begin{equation}
 \mbh=f_g{\sigma_T\over  m_p}{\sigma^4\over{\pi G^2}}=2\left(\frac{f_g}{0.1}\right)\sigmatwo^4\;  10^8M_\odot  \; , 
 \label{eq:fbalance}
 \end{equation}
  very close to the observed $\mbh-\sigma$ relation
$\mbht/\sigmatwo^{4.24\pm 0.41}=10^{0.12\pm 0.08}$ \citep{gult09}, where  $\mbht=\mbh/10^8\Msol$ 
and $\sigmatwo=\sigma/200 \kms$. 

These arguments do not take into account the lifetime of the  AGN.  The following consideration shows that black holes obeying the observed $\mbh -\sigma $ relation cannot  generate enough energy in radiation in order to drive the gas out of the protogalactic potential 
well by radiation pressure. 
The work, $L(r_e-r)/c$,
done by radiation pressure in moving the gas from $r$ to $r_e$ must be sufficient to bring the 
gas to the escape speed, $v_e(r_e)=\sqrt{-2\phi(r_e)}$, at $r_e$. Energy conservation  then demands  
\begin{equation}
\frac{L}{c}(r_e-r)>\frac{1}{2}  \mg v_e^2(r)\; ,
\end{equation}
 where $v_e(r)=\sqrt{-2\phi(r)}$ is the escape speed at $r$.
The  total energy radiated by the AGN during this process
 is $L \tau$ where $\tau =\int_{r}^{r_e} \dd r /v(r)<(r-r_e)/v_e(r)$. Therefore, 
the energy radiated by the AGN must be greater than 
\begin{equation}
E_m=L(r_e-r)/v_e(r)=\frac{c}{2}  \mg v_e(r)\; 
\end{equation}
For an isothermal sphere truncated at the virial radius $R_v$, we have 
$V_e^2(r)=4\sigma^2(1+\ln\frac{R_v}{r})$
so that
$ E_m>  \mg c \sigma . $
The total energy $\eta \mbh c^2$ ($\eta\sim 0.1$ is the efficiency factor) that could be extracted from the black hole 
must therefore satisfy,
\begin{equation}
\label{eq:Econd}
\eta \mbh c>\mg  \sigma \; . 
\end{equation}
as the condition for momentum-driven winds to be able to unbind galactic gas.
The virial  gas mass in a halo is 
$
\mg=7.3\times 10^{11}\sigmatwo^3 (\fg/0.1)/h(z)\; M_\odot\; ,
$
where $h(z)=H_0/H(z)$ and we have used $H_0=71 \kms /\Mpc$.
Therefore, 
\begin{equation}
\label{eq:minE}
\eta \mbh  c>5\times 10^8 \sigmatwo^4 (\fg/0.1)/h(z)\; M_\odot c\; .
\end{equation}
Using the observed $\mbh$-$\sigma$ relation, we find that the black hole  cannot unbind
the gas by momentum even for $\sigma=400\rm km/s$ at $z=2$. The shortfall is  about an order of magnitude at $z=0$.

Complementary arguments are recently given by \cite{Anderson2010}. These authors confirm the ubiquitous baryon fraction deficiency in galaxies. They compare baryon fractions in galaxies with varying bulge-to-disk-mass, and demonstrate that the presence of a supermassive black hole does not result in a reduced baryon fraction.
%
%

In figure \ref{fig1} we plot the ratio $0.1 \mbh c/(\mg \sigma)$ versus $\sigma$  from  the galaxy 
sample  in Table 5 of  \cite{gult09}.  As a  proxy  for $\mg $, the total gas mass to be ejected,  we conservatively use the V-band luminosities of galaxies in the sample  multiplied by a stellar mass-to-light ratio of 4.   For the majority of galaxies in this sample the maximum momentum $\eta \mbh c$ falls short of $\mg \sigma$ by  a factor of a few: half the sample falls short of the required threshold by a factor of at least 4,  a quarter by a factor of 10.  The figure also implies that the ratio 
depends weakly on  the
velocity dispersion.

Multiple scatterings can modulate the momentum delivered to the shell by the radiation but fail to resolve the momentum shortfall. Here is our reasoning.
In the scattering case, force balance implies that $$\rho g(r) =\kappa \rho {{L}\over{4\pi c r^2}},$$
where $\rho$ is the gas density. Assuming $g(r)$ is constant over the  shell, we
integrating this relation over the gas shell to give
$g M_g=\tau L/c,$ where $\tau$  is proportional to $\sigma$  in the cosmological setting. Assuming $L$ is proportional  to $M_{BH}$ and since $gM_g$ is proportional to $\sigma^4,$  this force balance relation implies that $M_{BH} \propto   \sigma^3.$   Modulation by multiple scattering is therefore inconsistent with observations  because $\tau$ scales  with $\sigma.$   

To obtain the correct relation  $M_{BH} \propto   \sigma^4,$ or possibly steeper,   we consider a photon delivering its momentum in a single encounter with a gas particle. This is achieved if the gas surrounding the black hole is mostly neutral, which is plausible if the cooling time is short in the central regions. Predominance of molecular gas in the nuclear regions is inferred indirectly from dust, CO and star formation observations. The optical depth for dust also scales with surface density \citep{Thompson05}.

\section{The case against energy-driven outflows}

\begin{figure*}
\epsscale{1.2}
 \plotone{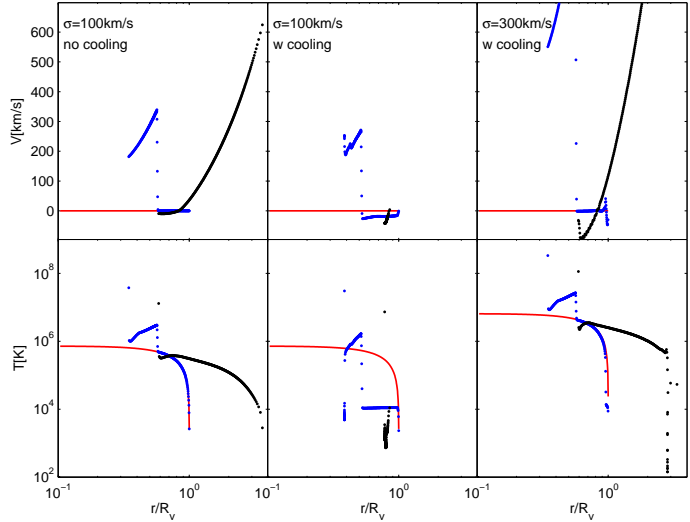}
\caption{The effect of radiative cooling on the ejection of gas by a central explosion in 
isothermal halos. Left, middle and right columns, respectively,  correspond to $\sigma=100\; \kms$ without cooling, 
$\sigma=100\; \kms$  with radiative cooling, and $\sigma=300\; \kms$ with cooling. 
Solid red lines represent the hydrostatic initial conditions. Blue dots  represent the systems at some intermediate time. Black 
dots correspond to the final time of the runs at 1.3 Gyr after the explosion. Note that this  1-d simulation does not describe fragmentation of shells. }
\label{fig2}
\end{figure*}

Energy-driven  outflows have been argued to  give the wrong scaling relation \citep{SR}    for  the observed $M_{BH} - \sigma$ relation, although a  recent reanalysis  favours the originally predicted  $M_{BH} - \sigma^5$ dependence \citep{graham}. However there is a more fundamental difficulty with energy-driven outflows. 
The gas  initially cools by Compton scattering with the radiation emitted by the AGN \citep{King03}. This cooling indeed is important in the central region but  lasts only as long as the AGN is active. { Radiative cooling plays an important role in the  AGN-galaxy interplay over  longer 
timescales.  Radiative cooling is extremely efficient in small halos ($\ltsim 5 .10^{11}\rm M_\odot$), where the cooling radius  can even exceed the virial radius  \citep{S77, RO77, WR78}. 
To explore the role of radiative cooling, we have simulated the feedback effects 
in spherical systems using a one-dimensional Lagrangian hydrodynamical code (see 
\cite{NusPo} for numerical details.) In these simulations, the dark matter is assumed to reside in 
a static isothermal spherical halo truncated at the virial radius, $\Rv$,  and the gravity of the gas is ignored. 
The gas is represented by 250 shells, which are equally spaced between $r=.1\Rv$ and $ \Rv$, and 
are  initially in hydrostatic equilibrium in the gravitational potential of the dark matter, with 
zero external pressure at $r=\Rv$. 
AGN feedback is introduced as heat added to the innermost shell  over a time-scale 
of $5\times 10^7$ yr, which is shorter than  the dynamical and the radiative cooling time-scales. 
The equilibrium radiative cooling curve with metallicity of one-third solar 
as given in \cite{DS} is adopted. 
Figure  \ref{fig2} shows the results from  the simulations for two halos with 
$\sigma=100\; \kms$  and $\sigma =300 \; \kms$. 
The explosion energy is taken to represent the AGN feedback, and is equal to the gas potential energy in absolute value. 
The integrated AGN energy delivered to the system is 
set equal to the initial potential energy of the gas, about $1.4\times {59} \; \rm ergs$ 
and $3.4\times {61} \; \rm ergs$ for $\sigma=100\; \kms $ and $300 \; \kms $, respectively. 
The results in the left panel (no cooling case) demonstrate that for $\sigma=100\; \kms$ without cooling, the gas is ejected from the system. 
For this smaller halo the radiated energy is so large 
that the total energy of the system at the final time is negative  and the gas falls back onto the halo, 
as seen in the middle panel.  Only for massive galaxies, 
$\sigma=300\; \kms $, right panel,  does cooling play a lesser role, so that the 
 feedback actually manages to unbind the gas out of the halo. 
 For the effect of the cooling to be negligible, a self-regulating 
 mechanism must operate so that the  gas is rapidly  ejected before cooling becomes important.
 But such a contrived mechanism will likely produce 
 a complicated functional form for the $\mbh-\sigma $ relation which 
 should reflect the dependence of the cooling curve on $\sigma$ and the details of how the energy is 
 ejected. 
}
Our discussion of the relevance of cooling demonstrates that momentum-driven outflows 
\citep{Fabian99, King2005}  dominate over all mass scales of interest and, as we have shown,   possibly give the correct scaling.
Such flows are inevitable, as  radiative cooling  dominates over a longer time scale, especially in smaller systems. However, they cannot  account for  the observed normalisation of the
$\mbh-\sigma $ relation.

\section{The role of star formation  triggering by AGN momentum-driven winds} 

If AGN and  SNe fail, admittedly for different reasons, to drive the required outflows, then we now argue that the combination may provide an ideal  solution. For a black hole emitting at the Eddington luminosity, force balance yields
$\mbh\sim \sigma^4$,  close to the observed scaling (see eq.~\ref{eq:fbalance}).
The problem as we have seen is that the black hole cannot supply sufficient energy 
to suppress its own growth by expelling gas by radiation pressure, i.e. the black hole does
not operate for long enough. The shortfall is a factor of a few.  This conclusion is also sustained by recent observational data \citep[e.g.,][]{DUNN}.
We propose here that  a radiation  momentum-driven wind triggers 
a starburst which joins forces with radiation to drive gas   out of the protogalaxy.
In this way,  radiation momentum-driven winds, under certain physical assumptions,  could  
yield a near  $ \mbh\sim \sigma^4$ scaling
although they are not solely responsible for gas expulsion. 

If the triggered starburst is to aid in unbinding the gas by heating it above the halo's virial 
temperature ,then 
$N_{SN}E_{SN}\approx 0.5 M_g \sigma^2\; , $
where $E_{SN} \approx 10^{51} \rm ergs$ is the mechanical energy release per SNII and $N_{SN}$ is the number of SNII in the starburst. Further,  the whole process of gas removal 
 has to be short enough  so that radiative cooling is 
not important. 
Since $M_g \sim \sigma^3$ and  {if we assume $N_{SN}\sim \mbh$}, we get $\mbh \sim \sigma^5$.
This    corresponds to  the observed   scaling in the energy-driven wind case and allows an acceptable normalisation.  Alternatively, to achieve a $ \sigma^4$ scaling, our preferred model invokes considerations of momentum balance. 
The starburst must supply the ``missing momentum" of a factor of a few of the black hole's $\eta \mbh c$. 
The momentum boost works as follows. A supernova remnant  (SNR) conserves energy until the swept-up shell mass decelerates to a velocity below 
$v_s\approx 400 \rm km/s$ \citep{CM},
below which momentum is approximately conserved (in a uniform medium). The resulting inefficiency of SNR energy input is of order $v_{cl}/v_{SN}$, where $v_{cl}$ is the interstellar cloud velocity dispersion
and $v_{SN}\approx E_{SN}/[m_{SN}v_s],$ where $E_{SN}\approx 1.5\times 10^{51}\rm ergs$ is the initial SN explosion energy
and $m_{SN} \approx 150 \rm M_\odot $ (for a standard  IMF) is the mass in stars required to form a supernova of type II. This inefficiency, of order a percent,  is confirmed for more realistic conditions by numerical simulations of SN-driven galactic winds \citep{dubois}.

The momentum input is 
$E_{SN}/v_s$
  per supernova, and momentum balance for 
  $N_{SN}$ 
  supernovae gives
  \begin{equation}
   N_{SN}E_{SN}/v_s=M_g\sigma\; .
  \label{eq:sbm}
  \end{equation}
  Since the gas mass $M_g\simpropto \sigma^3$ this yields a boost satisfying the desired dependence on $\sigma$. 
 The momentum boost $N_{SN}E_{SN}/v_s$ must be larger than $\eta \mbh c$ by a factor of a few. 
 To obtain this,  we assume that the black hole growth rate is equal to the Eddington-limited accretion rate and that the star formation luminosity is equal to the Eddington luminosity. This is  a reasonable approximation at zeroth order: in fact \cite{Netzer09} shows that there is a small tilt in the logarithmic relation for the star formation luminosity dependence on the luminous AGN that concern us here, with 
 $$L_\ast \simpropto L_{bol}^{0.9},$$ where the bolometric luminosity is the sum of star formation and AGN luminosity. Our  neglect of  this tilt  
leaves the boost  factor independent of mass and preserves the $\sigma^4$ dependence of the $\mbh, \sigma$ correlation. 
Therefore, we simply assume that the total star formation energy is $E_\ast =f_\ast \eta \mbh c^2$, where $f_\ast$  is a factor of order unity. 
The stellar mass  associated with $E_\ast$ is  $M_\ast=E_\ast/(\epsilon_{nuc} c^2)=f_\ast (\eta/0.1) 10^{10}\mbht $, where   $\epsilon_{nuc} \approx 10^{-3}-10^{-4}$   is the thermonuclear burning energy efficiency for a massive star.  
The associated mechanical energy produced by 
SNII in the starburst is 
\begin{equation}
N_{SN} E_{SN}/vs = \frac{ f_\ast E_{\ast}  E_{SN} }{\epsilon_{nuc} m_{SN}  c^2 v_s}=f_{_{boost}} \eta \mbh c \; .
\end{equation}
  Also   $N_{SN}=  M_\ast  / m_{SN},$ where $m_{SN}\approx 100\rm M_\odot$ is the mass  in stars formed per SNII and is weakly IMF-dependent.
The boost factor,
\begin{equation}
f_{_{boost}}=\frac{f_\ast E_{SN}  }{\epsilon_{nuc} m_{SN}  c v_s } 
\end{equation} 
amounts to an order of magnitude for $f_\ast=1$,  $E_{SN}=10^{51}\; \rm ergs$, $\epsilon_{nuc} =3.10^{-4}$, $m_{SN}=150 M_\odot$ (expected for  e.g. a Chabrier IMF), and $v_s=400\kms$. 
Of course the question remaining is why the boost factor should be only weakly dependent on black hole  mass. This requires more than a universal mass function of gas clouds since the AGN outflow pressure enters. But this might work, see eq.~85 in \cite{SN09}: the AGN-driven supersonic turbulence velocity dispersion is found to depend only logarithmically on AGN properties, as also does  the porosity  which controls turbulent pressure.

We now discuss the implications of our feedback models for  further applications, notably  to satellite abundances, intermediate mass black holes,  and the baryon fraction.

\section{Feedback, satellite abundances and baryonic fraction} 

AGN-triggered  preheating is a plausible mechanism for reducing satellite abundances in galaxy groups or around massive early-type galaxies.
\cite{Koposov} demonstrated that SN feedback  plus reionization accounts for the luminosity function of the MWG and M31 dwarfs below 
$10^8 M_\odot$. However, there are  {more} intermediate mass  galaxies(namely LMC/SMC/M32/NGC205)  in the MWG and M31 halos  than are found in models tuned to fit the ultrafaint dwarf frequency. Models which fit the ultrafaint dwarfs are so efficient that they underproduce the massive dwarfs/intermediate mass galaxies. Conversely models tuned to fit the  massive dwarfs  have excessively inefficient  feedback and  overproduce   the numbers of ultrafaint dwarfs. This problem seems to be common to all SAMs. 
\cite{Smith} confirms that SAMs fail to resolve this problem for the abundance of intermediate mass galaxies.
This data set confirms that  massive galaxies are overproduced in the models.
\cite{Liu} also finds that all SAMs overpredict the number of satellites by  at least a factor of two in the
mass range $10^9-10^{10}\rm M_\odot$.

AGNs are commonly introduced at early epochs to account for the black hole--bulge correlation via the quasar feedback mode at early epochs.
At late epochs, the  AGN radio mode  inhibits cooling of the dilute gas resulting from the earlier feedback process, keeps the gas hot, resolves  the galaxy luminosity function bright end problem, and accounts for the red colours of massive early type galaxies.
We point out here that AGN feedback in the radio-quiet mode may also account for the suppression in numbers of dwarf satellite galaxies.  Feedback from AGN in the host galaxies preheats the halo gas  that otherwise would be captured by satellites.
{ However,  suppression of the formation of intermediate mass satellites of} the MW, and more generally, late-type galaxies with small bulges   {may not be efficient because of the low masses of their  central black holes.}  We have no solution to this problem, more generally associated with the large observed frequency of essentially bulge-less thin disks, other than to suggest patchy accretion of cold gas must play a role in thin disk formation at late epochs (but see \cite{PN}).
Once the potential well of a massive galaxy has developed, SN do not eject gas, although they may drive interstellar turbulence and fountains. Gravitational heating does not work en route to forming the potential. All that is left is vigorous activity in the MW assembly stage. This phase may plausibly involve feedback from intermediate mass black holes, which are believed, at least by some, to be ubiquitous.

The hypothesis that intermediate mass black holes are formed generically during the hierarchical build-up of galaxies may 
possibly provide a radical solution to the { baryon fraction}
  problem via the momentum-driven outflows that we are invoking. 
Theoretical arguments suggest that one pathway towards 
building up the central SMBH is via mergers of intermediate mass black holes (IMBH) during the hierarchical merging evolution of dark matter halos. It is assumed that substructures develop IMBH
at early epochs, contemporaneously with first star formation. 
Simultaneously, another major problem is resolved, that of the baryon fraction, via preheating or ejection. This is seen to be low in low mass and in massive galaxies (McGaugh et all 2009). If satellites form in a secondary manner, preheating reduces  their baryon fraction. If satellites formed first, the quasar mode will sweep the gas out of the galaxy. This is achieved by a combination of  AGN outflow momentum plus induced SN feedback. The gas subsequently stays out via the quiet mode of AGN feedback from the central host and other active galaxies.

Within the standard paradigm, the observed baryon fraction  of the MWG with its small black hole is only explained if there are sufficient IMBH in the satellites to drive out the baryons at early epochs (see \cite{KNP10} for  an alternative non-standard explanations). For galaxies with massive spheroids, the central BH strips the satellites, reducing the number of IMBH, but provides enough feedback to eject the baryons.

Baryons   must be ejected in order to begin with the primordial baryon abundance, from galaxies  of all mass scales, as well as 
galaxy groups. Only at clusters scales is the baryon fraction convergent. For dwarfs this is not a problem, but for typical galaxies, such as the MW,  SN feedback cannot be responsible for baryon loss. Rather the baryons are recycled via the halo. AGNs provide the only energy source capable of accounting for baryon ejection.

\section{Discussion}

The primary aim of our   paper is to highlight the scaling relation  normalisation problem for AGN feedback,  and to  propose a possible solution involving AGN-triggered star formation. Positive feedback may have important ramifications for star formation at high redshift, and  is inevitably followed by gas outflows driven by both AGN and supernovae, along with concomitant quenching of star formation.

From the data plotted in figure 1,  the momentum boosting by the starburst is a factor of a few. This will naturally yield the dispersion in the relations given the nature of the boost, e.g. by BH outflow  triggering of SNII.  The points that lie low  in the momentum condition had a larger boost, and this would  lead to a prediction that the residuals in  $\mbh c$ vs $M_g\sigma$ should anticorrelate with SNII tracers in chemical evolution, e.g. the  bulge $\alpha/Fe$.

Small galaxies which formed stars before host galaxy AGN onset will survive.
 They should be seen as a bump in the galaxy luminosity function (GLF), analogously to what is seen in the MW \citep{Koposov} and in the  K-band GLF \citep{Smith}.
These galaxies are distinguishable by being older and more metal-poor than their AGN-modulated successors which are primarily either low mass satellites or  massive early-type galaxies.
For the MW, the failure of the \cite{Koposov} model tuned to the numerous ultrafaint dwarfs to account for the admittedly sparse numbers of massive dwarfs is consistent with the lack of a large BH 
in our AGN feedback model. Feedback from IMBH can resolve this problem. We have suggested that the IMBH in $\omega$Cen may be an example of a population of  halo IMBH that could have provided the additional feedback needed to both allow the LMC and similar dwarfs  to form and not  simultaneously overproduce the faint dwarfs. Such IMBH could easily, during an active accretion phase, have produced enough momentum to have swept the residual gas out of the outer halo. 

Globular clusters are plausibly the most visible surviving component of the first generation of substructure. That they might have a direct connection to IMBH is weakly supported  by the possibility that  one of the most massive globular clusters, $\omega$Cen,  might contain an IMBH. Another hint of 
a connection with globular clusters may be present in the apparent correlation  between black hole 
mass and mass of the host galaxy globular cluster system \citep{Spitler}. A variation on this relation has 
recently been found that relates black hole mass to the number of globular clusters \citep{Burkert2010}.
Numerical simulations find that the SMBH-$\sigma$ scaling relation can be preserved by hierarchical mergers of IMBH \citep{Burkert}. This lends support to the possibility that globular clusters may serve as a proxy both for IMBH and for dwarf galaxies, and therefore provide a possible witness to 
the required baryonic cleansing role of satellites by IMBH  in our model.

 \acknowledgements{We thank Noam Soker and Avi Loeb  for valuable comments. This work was partially supported by the ISRAEL SCIENCE FOUNDATION (grant No. 203/09), the Asher Space Research Institute and the WINNIPEG RESEARCH FUND.}

\bibliography{bibs}
\end{document}